\documentclass[%
 reprint,
 amsmath,amssymb,
 aps, showkeys,
pra,
]{revtex4-1}

\usepackage{graphicx}
\usepackage{dcolumn}
\usepackage{bm}
\usepackage[caption=false]{subfig}

\begin{document}

\preprint{APS/123-QED}

\title{High efficiency measurement of all orbital angular momentum modes in a light beam}

\author{Haad Yaqub Rathore}
\author{Mumtaz Sheikh}
\email{msheikh@ur.rochester.edu}
\altaffiliation[Also at ]{Institute of Optics, University of Rochester, Rochester NY 14627, USA.}
\author{Usman Javid}%
\author{Hamza Ahmed}
\author{Syed Azer Reza}
\affiliation{%
 Department of Physics, School of Science and Engineering, Lahore University of Management Sciences, DHA, Lahore Cantt., Pakistan 54792
}%

\date{\today}

\begin{abstract}
We present an experimental demonstration of a Laguerre-Gauss (LG) spectrum measurement technique using variable focus lenses that is able to measure the strengths of all modes present in an unknown, incoming light beam with the highest possible efficiency. The experiment modifies the classical projective, phase flattening technique by including a variable sized pinhole and a two electronic lens variable imaging system that is tuned for each mode to give the highest possible detection efficiency irrespective of the beam waist of LG mode chosen for the projection/decomposition. The modified experiment preserves the orthogonality between the modes with only a 4 \% cross-talk so that superposition states may also be detected efficiently. Our experiment results show efficient detection of OAM vortex beams with topological charge, $l$, values ranging from 0 to 4 with various different beam waists chosen for the decomposition.
\end{abstract}

\maketitle

\section*{Introduction}
Light beams with a helical phase structure can carry a definite value of orbital angular momentum (OAM) \cite{Allen:92}. Laguerre-Gauss (LG) laser modes are examples of such beams with a helical phase structure. Such beams can, in principle, carry an infinite amount of information since there are an infinite number of possible OAM states of a photon \cite{PhysRevLett.88.013601,Gibson:04}. Naturally, there has been a lot of interest over the years in using such beams for both classical \cite{Wang:12,Bozinovic:13} and quantum \cite{molina2007twisted, mirhosseini2015high} communication. For such applications, it is critical to have an efficient mechanism for measurement of the OAM spectrum of such beams. Various methods have been proposed for the generation and detection of such beams which have their respective advantages and disadvantages \cite{Mair:01,Leach:02,Vasnetsov:03,Karimi:09,PhysRevLett.105.153601,Lavery:11,Giovanni:12,mirhosseini2013efficient}. An excellent discussion of these methods is contained in Ref. \cite{mirhosseini2013efficient}.

We use the projective phase flattening appraoch, where an incoming unknown beam is projected on to conjugate LG modes, one mode at a time \cite{Mair:01}. If there is a mode-match with the incoming beam, the resulting beam when Fourier-transformed produces a central bright with a ringed intensity pattern which can couple into a single mode fiber (SMF). In essence, the unknown incoming beam is decomposed into orthogonal LG modes and subsequently its OAM content can be measured. 

It was shown in \cite{Qassim:14} that the detection efficiencies are mode dependent and that for a particular mode, the detection efficiency varies with the beam wasit, $w_0$, chosen for the decomposition. It was also shown that the maximum possible detection efficiency, corresponding to a particular value of $w_0$, decreased with mode order. Consequently, this rules out the use of higher order modes in spatial mode division multiplexed systems, thus severely limiting the use of OAM beams for practical high-speed optical communication applications. The limitations of the projective phase-flattening approach are further discussed in Ref. \cite{Qassim:14}. 

Recently, we proposed \cite{Sheikh:15,Sheikh:16} a method to overcome this particular limitation by modifying the classical method through the introduction of two variable focus lenses and a variable sized pinhole. The method works by dynamically correcting the phase curvature and the size of the beam coupling into the SMF based on the mode under consideration and the $w_0$, chosen for the decomposition. The simulation results demonstrated that the detection efficiencies for all modes are simultaneously the maximum possible irrespective of the choice of $w_0$. In this paper, we design and implement an experiment based on the schematic idea of Ref. \cite{Sheikh:16} and provide proof-of-concept experimental results to demonstrate the validity of the proposed scheme.

\section*{Description of the experiment}
An LG mode is characterized by two indices $l$ and $p$, which represent the azimuthal number and radial index of the mode respectively. These LG modes are mutually orthogonal and form a complete set of solutions to the paraxial wave equation. The mode can be mathematically represented by Eq. \ref{eq:eq1} at the pupil:
\begin{equation}
LG_{p, l}(r,\phi) = \sqrt{\frac{2^{|l|+1}p!}{\pi w_0^2 (p+|l|)!}}
\left(\frac{r}{w_0}\right)^{|l|} e^{-\frac{r^2}{w_0^2}} L_p^{|l|} \left(\frac{2r^2}{w_0^2}\right)
e^{-il\phi}
\label{eq:eq1}
\end{equation}
\noindent
where $r$ is the radial coordinate, $\phi$ is the azimuthal angle, $w_0$ is the beam waist radius at the pupil and  $L_p^{|l|}(.)$ is the generalized Laguerre polynomial.

\begin{figure}[htbp]
\centering
\includegraphics[width=\linewidth]{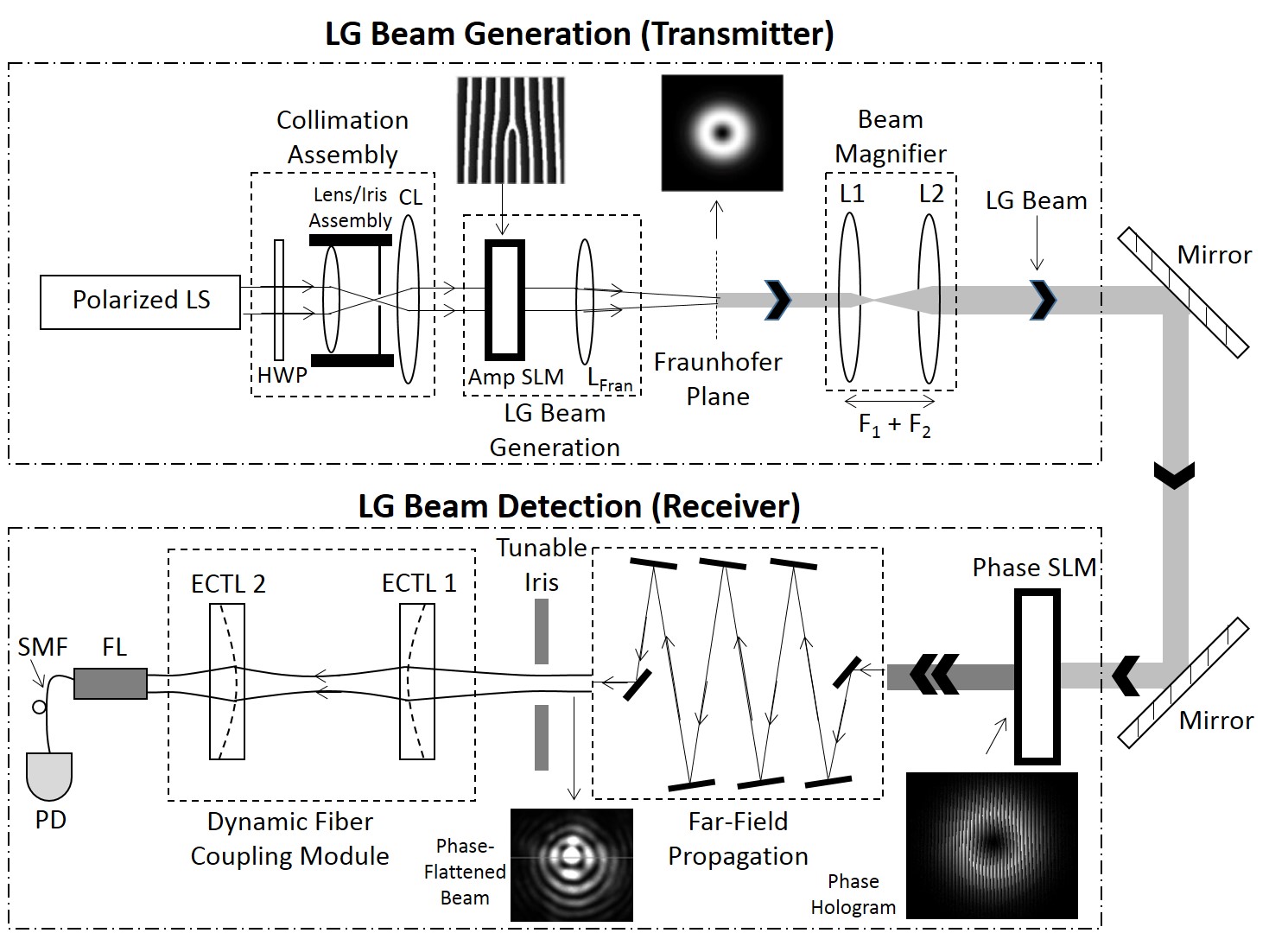}
\caption{Experiment to measure the LG spectrum of an unknown, incoming beam.}
\label{fig:fig1}
\end{figure}

Figure \ref{fig:fig1}
shows the proof-of-concept experiment set up in the lab to demonstrate the efficacy of our technique proposed in Ref. \cite{Sheikh:16}. In the first part of the experiment labeled "Transmitter" in Fig. \ref{fig:fig1}, an OAM vortex beam with topological charge $l$ is prepared via the classical approach using a forked grating pattern set on an amplitude spatial light modulator (SLM) \cite{Yu:1992,Gibson:04,Mirhosseini:13}. The prepared beam is a superposition of LG modes with different radial indices $p$ but with the same azimuthal number $l$. 

Light from a polarized laser source (LS) is collimated via a spatial filter system consisting of an aspheric lens, a pinhole and a collimating lens before it is made incident on the amplitude SLM with a forked grating pattern corresponding to some specific topological charge, $l$. A Fourier lens, $L_{fran}$ delivers the Fourier transform of the field at the amplitude SLM to generate the desired OAM vortex beam at the Fourier plane in the first diffraction order. Next, this field is magnified using a two-lens magnification system, L1 and L2, before it is ready to be detected by the second part of the experiment labeled "Receiver" in Fig. \ref{fig:fig1}.
The detector has the phase SLM as its main component on which a conjugate LG mode is projected which cancels the helical phase of the incoming beam if the mode matches the $l$-value of the beam. In the far-field of the phase SLM, this gives a central bright Gaussian-like spot with a ringed intensity pattern, as shown in Figure \ref{fig:fig1}. This can be made to couple efficienctly into a single mode fiber (SMF), as will be described shortly. In contrast, if there is a mode mismatch, then a central dark spot is observed in the far-field which would not couple into an SMF exactly like the classical phase-flattening approach \cite{Mair:01}. Note that we employ a long physical propagation distance to get to the Fraunhofer region as opposed to using a lens as is typically used. This is important for improving the sensitivity of our subsequent dynamic fiber coupling module \cite{Qasim:15} as explained later. The phase-flattened central bright Gaussian-like spot is first passed through a variable pinhole (an adjustable iris in this case) and then through two electronically controlled tunable lenses (ECTLs) before coupling into the SMF. The variable pinhole is used to select only the central bright Gaussian-like part of the beam so that it remains Gaussian on passing through the two ECTLs. The two ECTLs are used to match the amplitude and phase profile of the incoming beam to the fundamental mode of the SMF. As shown in Ref. \cite{Sheikh:16}, for achieving the maximum possible detection efficiency, the size of the variable pinhole and the ECTL focal lengths only depend on the mode under consideration and the beam waist, $w_0$, of the LG conjugate mode chosen for the decomposition. Our experimental results show that this maximum possible efficiency can be achieved for each mode irrespective of the beam waist, $w_0$, chosen for the decomposition by tuning the focal lengths of the ECTLs and varying the size of the pinhole.

\section*{Experiment results}
Fig. \ref{fig:fig1} experiment is set up in the lab using a MellesGriot 5 mW, 633 nm HeNe laser source. The amplitude SLM used to generate the test beam is Holoeye LC-2002 liquid crystal device. A forked diffraction grating hologram is generated using the MATLAB code from Ref. \cite{Galvezcode}. In terms of the LG modes, the OAM vortex modes generated by passing light through the forked diffraction grating are not pure for the radial number (p) but are pure for the azimuthal number or $l$-value of the LG beam mode. The desired beam mode is produced in the first order of the diffraction pattern in the Fourier plane of the Amplitude SLM using the lens labeled $L_{fran}$ (focal length = 50 cm). Note that since the HOLOEYE amplitude SLM is a pixelated device, thus it forms a 2-D diffraction grating on top of the forked diffraction grating. Diffraction orders produced along the $x$-axis are due to two sources, the pixels of the SLM and the forked grating. We have to be careful to select the zeroth order of the SLM’s grating and the first order of the forked grating. Such a double diffraction grating leaves only a small fraction of the incident power in our desired beam.

\begin{figure}[htbp]
\centering
\includegraphics[width=0.6\linewidth]{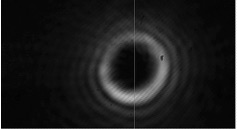}
\caption{Prepared OAM vortex beam for $l=2$.}
\label{fig:fig2}
\end{figure}

Since our next task was to cancel the helical phase of the generated LG mode using a reflective phase SLM and because the generated LG mode has a small beam radius (at the Fraunhoufer plane), we employ a classic beam expander with a 15-fold magnification using two lenses L1 and L2 with focal lengths $f_1=5$ cm and $f_2=75$ cm. The resulting spatial mode at the Fourier plane of the lens $L_{fran}$ is then projected onto the phase SLM. The prepared test beam for $l=2$, as seen on a CCD placed at the SLM plane, is shown in Fig. \ref{fig:fig2}. 

\begin{figure}[htbp]
\centering
\includegraphics[width=0.6\linewidth]{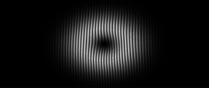}
\caption{Computer-generated hologram for the conjugate mode $p=0$, $l=-2$.}
\label{fig:fig3}
\end{figure}

\begin{figure}[htbp]
\centering
\subfloat[]{\label{fig:fig4a}%
  \includegraphics[width=0.4\linewidth]{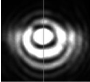}%
}
\subfloat[]{\label{fig:fig4b}%
  \includegraphics[width=0.4\linewidth]{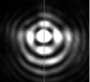}%
}\\
\subfloat[]{\label{fig:fig4c}%
  \includegraphics[width=0.4\linewidth]{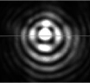}%
}
\caption{Phase-flattened beam observed at the Fourier plane of a lens placed in front of the phase SLM. (a) $l=3$ (b) $l=5$ (c) $l=6$.}
\label{fig:fig4}
\end{figure}

The phase SLM used is a Holoeye PLUTO-VIS reflective liquid crystal device. Since the device is polarization sensitive, therefore a half-wave plate placed in the incoming beam path is used to optimize the first diffraction order from the phase SLM. The holographic kinoform for the phase SLM was generated using our own MATLAB code. We used intensity masking and implemented it following the guidelines of \cite{Bolduc:13} with a blazed grating superposed on the intensity-masked kinoform. The computer-generated hologram for the conjugate LG mode $p = 0$, $l = -2$ is shown in Fig. \ref{fig:fig3}.
Two mirrors mounted on tip-tilt stages were used to fold the incoming beam and direct it onto the SLM screen. The phase flattening occurs if the center of the vortex in the incoming beam exactly coincides with the phase vortex in the hologram. Also it is important for better results that the incident angle of the beam directed onto the phase SLM screen is small. In our case, we ensured it to be roughly 7 degrees. Phase flattening only occurs in the first diffraction order of the blazed grating. Like the amplitude SLM, the phase SLM is also a pixelated device (acts as a 2-D grating), so we observe two different sets of diffraction orders. The resulting phase-flattened LG modes, for different values of OAM, observed in the Fourier plane of the phase SLM are shown in \ref{fig:fig4}. Here we have used a lens of focal length 100 cm to Fourier transform the field at the phase SLM. It is worth emphasizing that in all of the images shown in Fig. \ref{fig:fig4}, the $l$-value of the incoming beam matches that of the conjugate LG mode projected on the phase SLM. Notice that as the $l$-value of the beam is increased, the size of the central bright spot decreases.

\begin{figure}[htbp]
\centering
\subfloat[]{\label{fig:fig6a}%
  \includegraphics[width=0.4\linewidth]{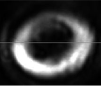}%
}
\subfloat[]{\label{fig:fig6b}%
  \includegraphics[width=0.4\linewidth]{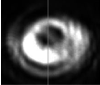}%
}
\caption{The case of mode-mismatch at the phase SLM observed at the Fourier plane of a lens placed in front of the phase SLM. The incoming test beam had an OAM of 2$\hbar$ per photon. (a) Phase SLM hologram set to $l=1$ (b) Phase SLM hologram set to $l=-1$.}
\label{fig:fig6}
\end{figure}

If the hologram corresponding to a different $l$-value was set on the phase SLM, we observe that a central bright spot does not appear as the helical phase is not completely flattened. Instead, a central dark spot is obtained. Our results for two different cases of mode mismatch are shown in \ref{fig:fig6}. As before, we Fourier transformed the plane of the phase SLM using a lens of focal length 100 cm to obtain these spatial distributions. This is a classical result i.e. the greater the mismatch, the greater in size is the central vortex.

Next, instead of a Fourier lens employed after the phase SLM, we use a 750 cm free-space propagation to get to the Fraunhofer regime.  We notice that if we use a lens (even a 100 cm one) to Fourier Transform the field at the plane of the phase SLM, the intensity distribution evolves rapidly after the Fourier plane. This rapid evolution is undesirable as it places a tighter limit on the tuning sensitivity and the range of the ECTLs compared to a free-space propagation for efficient coupling of light into the SMF.

\begin{figure}[htbp]
\centering
\subfloat[]{\label{fig:fig7a}%
  \includegraphics[width=0.4\linewidth]{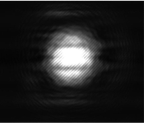}%
}
\subfloat[]{\label{fig:fig7b}%
  \includegraphics[width=0.4\linewidth]{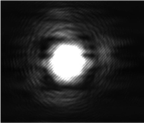}%
}\\
\subfloat[]{\label{fig:fig7c}%
  \includegraphics[width=0.4\linewidth]{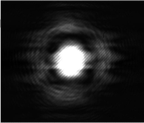}%
}
\subfloat[]{\label{fig:fig7d}%
  \includegraphics[width=0.4\linewidth]{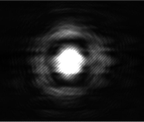}%
}\\
\subfloat[]{\label{fig:fig7e}%
  \includegraphics[width=0.4\linewidth]{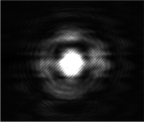}%
}
\caption{Phase-flattened beam observed in the Fraunhofer plane of the phase SLM for fixed $w_0 = 1.30$ mm and (a) $l=0$ (b) $l=1$ (c) $l=2$ (d) $l=3$ (e) $l=4$.}
\label{fig:fig7}
\end{figure}

\begin{figure}[htbp]
\centering
\includegraphics[width=0.6\linewidth]{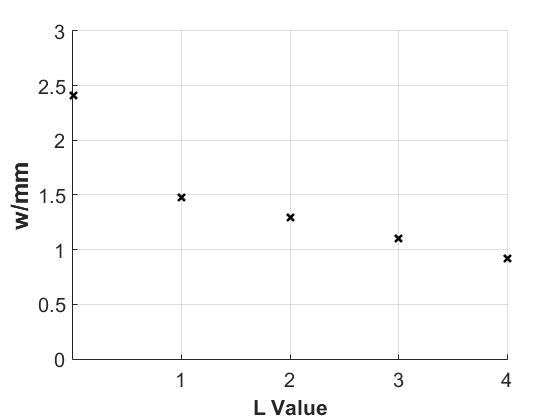}
\caption{Variation of the size of the central bright spot with the $l$-value of the beam.}
\label{fig:fig75}
\end{figure}

As the $l$-value in the beam is varied, the spatial mode profile of the phase flattened beam changes. In particular, the diameter of the central bright spot decreases as $l$ is increased. The spatial profiles of the phase flattened beam, imaged at a distance of 750 cm from the phase SLM, are shown in Fig. \ref{fig:fig7} for the first five values of $l$.  
These modes are obtained after successfully phase flattening the LG mode by applying the appropriate hologram on the phase SLM with $w_0 = 1.30$ mm. Fig. \ref{fig:fig75} shows the observed variation of the size of the central bright spot with the $l$-value of the beam. Note that the central bright spot is the largest for $l = 0$ while is the smallest for $l = 4$ in accordance with the simulations in \cite{Sheikh:16}, Fig. 1.

\begin{figure}[htbp]
\centering
\subfloat[]{\label{fig:fig8a}%
  \includegraphics[width=0.4\linewidth]{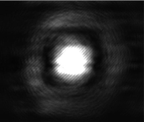}%
}
\subfloat[]{\label{fig:fig8b}%
  \includegraphics[width=0.4\linewidth]{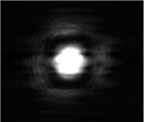}%
}\\
\subfloat[]{\label{fig:fig8c}%
  \includegraphics[width=0.4\linewidth]{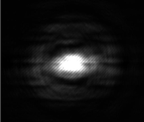}%
}
\caption{Phase-flattened beam observed in the Fraunhofer plane of the phase SLM for fixed $l=2$ and (a) $w_0=1.0$ mm (b) $w_0=1.3$ mm (c) $w_0=1.6$ mm.}
\label{fig:fig8}
\end{figure}

On the other hand, if we keep the $l$-value fixed and vary $w_0$ for the conjugate mode at the phase SLM, we find an inverse relation of the size of the central bright spot with $w_0$. This was precisely simulated in Ref. \cite{Sheikh:16}, Fig. 2. The observed spot in the Fraunhofer regime of the phase SLM for $l = 2$ is shown in Fig. \ref{fig:fig8}. 

\begin{figure}[htbp]
\centering
\subfloat[]{\label{fig:fig9a}%
  \includegraphics[width=0.4\linewidth]{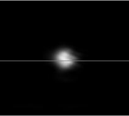}%
}
\subfloat[]{\label{fig:fig9b}%
  \includegraphics[width=0.32\linewidth]{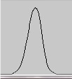}%
}\\
\caption{Phase-flattened beam truncated by the adjustable pinhole (iris) (a) observed on the CCD (b) beam cross-section.}
\label{fig:fig9}
\end{figure}

The phase-flattened beams shown in Figures \ref{fig:fig7} and \ref{fig:fig8} are truncated. We use an iris to remove the rings and make the spatial mode “Gaussian-like”. Since the size of the central bright spot is dependent on the value of $l$ and the value of $w_0$, hence the iris has to be adjusted every time even if one of the two parameters change. In an optimal setting, an amplitude SLM could have been used as the variable sized pinhole but in our case, since the beam spot is sufficiently large for all the modes and using the amplitude SLM results in significant losses due to diffraction, the iris works reasonably well. The intensity profile of a truncated beam ($l = 4$, $w_0 = 1.6$ mm) is shown in Fig. \ref{fig:fig9}.
Note that in Fig. \ref{fig:fig9b}, the intensity profile after truncation by the pinhole is Gaussian-like.

\begin{figure}[htbp]
\centering
\includegraphics[width=0.6\linewidth]{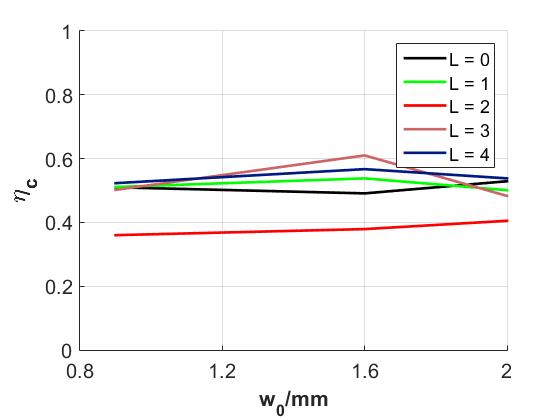}
\caption{Variation of the coupling efficiency, $\eta_c$ with the LG mode beam waist $w_0$ chosen for the decomposition.}
\label{fig:fig95}
\end{figure}

\begin{figure}[htbp]
\centering
\subfloat[]{\label{fig:fig10a}%
  \includegraphics[width=0.4\linewidth]{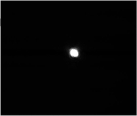}%
}
\subfloat[]{\label{fig:fig10b}%
  \includegraphics[width=0.4\linewidth]{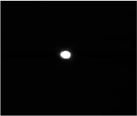}%
}\\
\subfloat[]{\label{fig:fig10c}%
  \includegraphics[width=0.4\linewidth]{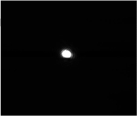}%
}
\subfloat[]{\label{fig:fig10d}%
  \includegraphics[width=0.4\linewidth]{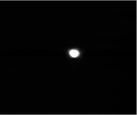}%
}\\
\subfloat[]{\label{fig:fig10e}%
  \includegraphics[width=0.4\linewidth]{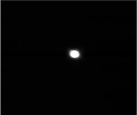}%
}
\caption{CCD images of the beam profiles just before the fiber coupler after tuning of the ECTLs for the modes (a) $l=0$ (b) $l=1$ (c) $l=2$ (d) $l=3$ (e) $l=4$.}
\label{fig:fig10}
\end{figure}

This Gaussian-like beam is now coupled into a SMF using two ECTLs and a fiber lens. The ECTLs used are Optotune liquid lenses Model EL10-30LD with active diameter 10 mm and a tuning range from 4 to 14 cm and a distance of 20 cm in between them. We tune the ECTLs to their optimal values so as to maximize the coupling efficiency into the SMF. The focal lengths of both the ECTLs are set to the ones calculated in Table 1 of Ref. \cite{Sheikh:16} as our starting point and are varied simultaneously to obtain the maximum coupling efficiency. The coupling efficiency, $\eta_c$ is measured by taking the ratio of the power being coupled into the SMF and the incident power right before the fiber lens. The intensity of light coupled into the SMF is observed to change as the focal lengths of the ECTLs are varied. Table \ref{tab:tab1} shows the results for five different values of $l$. For each value of $l$, we use three different values of $w_0$ (beam waist), a parameter set by the hologram on the Phase SLM.
\begin{table}[htbp]
\centering
\caption{\bf Observed parameters and detection efficiencies for different modes}
\begin{tabular}{ccccc}
\hline
Mode & $w_0$ & $f_1$  & $f_2$ & $\eta_c$ \\
(p,l) & (mm) & (cm) & (cm) & \\
\hline
$(0,0)$ & $0.9$ & 7.6 & 9.7 & 0.51 \\
$(0,0)$ & $1.6$ & 7.6 & 9.7 & 0.49 \\
$(0,0)$ & $2.0$ & 7.7 & 9.8 & 0.53 \\
$(0,1)$ & $0.9$ & 7.9 & 9.5 & 0.51 \\
$(0,1)$ & $1.6$ & 7.9 & 9.7 & 0.54 \\
$(0,1)$ & $2.0$ & 7.8 & 9.7 & 0.50 \\
$(0,2)$ & $0.9$ & 6.5 & 8.5 & 0.36 \\
$(0,2)$ & $1.6$ & 6.5 & 8.6 & 0.38 \\
$(0,2)$ & $2.0$ & 6.3 & 8.5 & 0.41 \\
$(0,3)$ & $0.9$ & 6.7 & 7.9 & 0.50 \\
$(0,3)$ & $1.6$ & 6.5 & 8.6 & 0.61 \\
$(0,3)$ & $2.0$ & 7.9 & 7.6 & 0.48 \\
$(0,4)$ & $0.9$ & 8.0 & 9.5 & 0.52 \\
$(0,4)$ & $1.6$ & 6.4 & 8.5 & 0.57 \\
$(0,4)$ & $2.0$ & 8.0 & 6.7 & 0.54 \\
\hline
\end{tabular}
\label{tab:tab1}
\end{table}
The ECTLs are controlled by varying the DC drive current. The actual observed focal lengths for the maximum possible coupling efficiency are slightly different from the simulated results because of hysteresis issues with liquid lens technology. For example, we have observed that depending on whether the drive current is being increased or decreased, a different focal length value of the ECTL is achieved.                  
The graph of $\eta_c$ against $w_0$ is shown in Fig. \ref{fig:fig95}.
The reason why our design achieves the highest possible coupling efficiency is that the tunable lenses mode match the radius of curvature and the beam spot to that of the fundamental mode of the SMF. Although, in principle, we can have coupling efficiencies approaching 100\% as simulated in Ref. \cite{Sheikh:16}, in practice, we just have these efficiencies approaching 50\% because of the difficulty in coupling light into a SMF especially at visible wavelengths with a very small fiber core. The same is also true for Ref. \cite{Qassim:14} where the coupling efficiencies achieved in the experiment were about half that of the simulated values. Having tuned the system parameters (the focal lengths of the ECTLs and the pinhole size), we image the beam profile right before the fiber collimator for each value of $l$ and unsurprisingly these intensity profiles are exactly the same (to within the limit of experimental accuracy) as shown in Fig. \ref{fig:fig10}.

It must be emphasized that had we not used the tunable lenses, the sizes obtained of the central bright spot, would have been differently sized like Fig. \ref{fig:fig7}. It can be clearly seen that in such a case, most of the power in the central bright spot would not have been coupled into the SMF because of the difference in the beam waists of the incoming beam and that of the fundamental mode of the SMF. Additionally, the coupling efficiencies would have been even lower due the presence of the quadratic phase associated with the propagation of the phase-flattened beam \cite{Sheikh:16}. Hence our experimental design not only matches the transverse field amplitude but also the phase curvature of the phase-flattened beam with the fiber mode.

For the mode mismatch case, we generate the $LG_{pl} = LG_{02}$ using the amplitude SLM and set the $LG_{03}$ ($w_0 = 1.60$ mm) on the phase SLM. The pinhole setting is kept exactly the same as the mode match case of $LG_{03}$ and the focal lengths used are $f_1 = 6.5$ cm and $f_2 = 8.6$ cm as given in Table \ref{tab:tab1}. We find that the coupling efficiency, $\eta_c$ drops to 0.091 in this case. The percentage power passing through the pinhole also drops because there is now a central dark spot in the beam (like Fig. \ref{fig:fig6}) and most of the power is in the radially outward portion of the beam. Similarly, setting the mode on the phase SLM to $LG_{04}$ ($w_0 = 1.60$ mm) and setting the pinhole exactly as for the mode match case of $LG_{04}$ gives us $\eta_c = 0.107$. The focal lengths, in this case, are fixed to those corresponding to $LG_{04}$ in Table \ref{tab:tab1}. According to our measurements, only about 23 \% of the incoming intensity passes through the pinhole so that the cross-talk is roughly only about 4 \%. As a final check, we generate $LG_{00}$ on the amplitude SLM which corresponds to merely a linear phase ramp applied on the amplitude SLM and set $LG_{02}$ on the phase SLM. The power coupled into the SMF drops down to 0.04 $\mu W$. On the contrary, for the mode match case, the power coupled into the SMF is 0.78 $\mu W$ (about 20 times as much). The robustness of our design to distinguish between different modes is apparent if we compare these efficiencies with the ones of the mode match case (see Table \ref{tab:tab1}).

\section*{Conclusion}
We have experimentally demonstrated a high efficiency LG spectrum measurement technique whereby different LG modes can be detected with the highest possible efficiency with low crosstalk between the modes. This is achieved by minimizing the coupling losses into a SMF using adaptive optics to vary the size and radius of curvature of the incoming beam. The actual modal detection efficiencies, although the maximum possible, are not the same and in any application, pre-biasing would be required.   

\bibliography{LGspectbib}

\end{document}